# Learning about Data, Algorithms, and Algorithmic Justice on TikTok in Personally Meaningful Ways

Luis Morales-Navarro (co-organizer), University of Pennsylvania, luismn@upenn.edu
Yasmin B. Kafai (co-organizer), University of Pennsylvania, kafai@upenn.edu
Ha Nguyen (co-organizer), Utah State University, ha.nguyen@usu.edu
Kayla DesPortes, New York University, kayla.desportes@nyu.edu
Ralph Vacca, Fordham University, rvacca2@fordham.edu
Camillia Matuk, New York University, cmatuk@nyu.edu
Megan Silander, Education Development Center, msilander@edc.org
Anna Amato, New York University, ada437@nyu.edu
Peter Woods, University of Nottingham, peter.woods@nottingham.ac.uk
Francisco Castro, New York University, francisco.castro@nyu.edu
Mia Shaw, New York University, mia.shaw@nyu.edu
Selin Akgun, Michigan State University, akgunsel@msu.edu
Christine Greenhow, Michigan State University, greenhow@msu.edu
Antero Garcia (discussant), Stanford University, Antero.Garcia@stanford.edu

**Abstract:** TikTok, a popular short video sharing application, emerged as the dominant social media platform for young people, with a pronounced influence on how young women and people of color interact online. The application has become a global space for youth to connect with each other, offering not only entertainment but also opportunities to engage with artificial intelligence/machine learning (AI/ML)-driven recommendations and create content using AI/M-powered tools, such as generative AI filters. This provides opportunities for youth to explore and question the inner workings of these systems, their implications, and even use them to advocate for causes they are passionate about. We present different perspectives on how youth may learn in personally meaningful ways when engaging with TikTok. We discuss how youth investigate how TikTok works (considering data and algorithms), take into account issues of ethics and algorithmic justice and use their understanding of the platform to advocate for change.

## Symposium overview

Over the past couple of years, TikTok has increased its popularity among youth in the United States with 67% of young people identifying as TikTok users and the majority of them using it on a daily basis (Pew Research Center, 2022). Notably, this trend is most prevalent among Black youth (81% of them reporting TikTok usage) and teenage girls (73%) (Pew Research Center, 2022). Research on social media and youth shows that these kinds of platforms may support young people to express themselves, develop their identities, build communities, and learn from each other (boyd, 2015; Jenkins et al., 2015). Education research highlights that social media can be productive for youth to engage in active learning, collaborate with others and build relationships with their communities (Greenhow et al., 2019). Yet, TikTok's pervasiveness in youths' lives, its use of AI/ML-driven recommendations, and emphasis on creative expression have changed the social media ecosystem as people become more aware of "the algorithm" (Bhandari & Bimo, 2022) and perceive it as more responsive than other platforms (Taylor & Choi, 2022). Indeed, as Karizat and colleagues (2021) argue, users build folk theories about how these algorithmic systems work in relation to their own identities and come up with plans to benefit from and resist expected algorithmic behaviors. At the same time, it is crucial for youths to become critical consumers and producers by engaging with ideas of algorithmic justice. Algorithmic justice considers that AI/ML algorithmic systems (in this case on TikTok) have implications on  individuals and communities that could perpetuate harm in unjust ways, disproportionately impacting vulnerable populations (Birhane, 2021). Rather than focusing on technical solutions, algorithmic justice places communities most likely impacted by AI/ML systems at the center of inquiries about knowledge, harm, and bias. In this symposium we investigate how youth learn about data, algorithms and algorithmic justice on TikTok.

We build on existing work in the Learning Sciences that investigates youths' meaning making and critical data practices (Nguyen & Parameswaran, 2023), youths' activism and political expression (Literat & Kligler-Vilenchik, 2023), and AI/ML ethics (Akgun & Greenhow, 2022) on TikTok. Presenters explore how youth engage in critical inquiry, which involves examining the implications and consequences of computing technologies (Morales-Navarro & Kafai, 2023), in personally meaningful ways, investigating the functionality and implications of data and algorithmic systems on TikTok as they create and consume content. Together we address the following questions: (1) How do youths perceive data and AI/ML-powered algorithms on TikTok? (2) What are different ways in which we can study youths' learning and interaction on this platform? (3) How can we leverage youths' experiences on TikTok to support the development of critical data and AI/ML literacies?

The invited works provide examples of studies on youth's evaluation of generative AI filters on TikTok and their views on potential harmful biases, organic learning practices on TikTok, teachers' perspectives on critical data literacy activities that build on students' experiences with TikTok, and critical AI literacy perspectives that center ethics. Participants contribute to Learning Sciences perspectives on critical computing education, data literacies, learning analytics, and AI/ML education. This session is organized in three sections: (1) the chairs will introduce the topic and participants (~10 min); (2) participants will have 10 minutes to share their work; (3) our discussant Antero García, whose recent work centers on designing education data for justice, will synthesize and reflect on findings (10 minutes) followed by a Q&A with audience and presenters (~15 min).

## Youth's everyday ideas about algorithmic justice in generative AI anime TikTok and Snapchat filters

Luis Morales-Navarro and Yasmin B. Kafai

Despite the increasing interest in fairness, accountability and justice research in computing, little attention has been paid to how children and youth perceive justice and fairness in relation to artificial intelligence applications they use in their everyday lives. Some of the studies that have investigated youth's ideas with regards to algorithmic justice have focused on high-stakes algorithmic systems engaging youth in discussions in relation to high-stakes policing surveillance technologies (e.g., Vakil & McKinney de Royston, 2022) and hypothetical robot interactions (e.g., Charisi et al., 2021) with less attention to everyday systems that youth interact with while looking for entertainment or connecting with friends. Yet, there is evidence that bias in algorithmic systems in popular applications such as TikTok and Snapchat can have negative impacts on youth's perceptions of themselves and on their own health (Burnell et al., 2022).

We investigated high school youth's ideas about algorithmic justice while conducting an informal audit of anime TikTok and Snapchat filters which tend to be popular among young people. Algorithm auditing is a method that involves "repeatedly querying an algorithm and observing its output in order to draw conclusions about the algorithm's opaque inner workings and possible external impact" (Metaxa et al., 2021) that has become popular in algorithmic justice and fairness research. Recently several studies have investigated how everyday non-expert users can find harmful algorithmic behaviors through everyday auditing (Shen et al., 2021; DeVos, 2022). Shen and colleagues (2021) define everyday algorithm auditing as "a process in which users detect, understand, and interrogate problematic machine behaviors" while interacting with algorithmic systems. Algorithm audits conducted by researchers on how these filters are trained and how they work demonstrate that the filters tend to reinforce harmful gender and racial biases (Jain et al., 2022). At the same time, informal audits shared on social media document how these filters do not work well on dark skinned and feminine-presenting people. Often, users can identify issues of algorithmic injustice that may not be detected through expert or centralized audits as these may be dependent on the context in which the user interacts with the system. While conducting everyday audits users have detected anticipated and potential algorithmic biases, developed theories and hypotheses that they tested to explain and understand how the algorithmic systems they use worked, and advocated for remediation, identifying plausible ways to address the issues identified (DeVos, 2022).

In computing education research audits have been discussed as productive opportunities for critical inquiry by having learners to investigate limitations and implications of computing applications (Morales-Navarro & Kafai, 2023). For example, Walker and colleagues (2022) adapted Buolamwini's (2022) ideas about "evocative audits" into a learning activity in which young African American students tested applications and used art to reflect on the harm that algorithms may inflict on their communities. Algorithm audits conducted by researchers on how these filters are trained and how they work demonstrate that the filters tend to reinforce harmful gender and racial biases (Jain et al., 2022). At the same time, informal audits shared on social media document how these filters do not work well on dark skinned and feminine-presenting people. We conducted a two-hour long workshop with 18 teenagers (ages 14-15) during which they conducted an informal algorithm audit in pairs. First, they tested the filters on each other, then they watched a video of different people using the filter, following they tested the filters against a set of pictures of popular artists, and finally they discussed and shared their findings with the larger group. Data was coded inductively looking at common themes across the data (Braun & Clarke, 2012).

We expected that youth would encounter cases in which the algorithms would not work for everyone, and that these cases would probe their ideas about algorithmic justice. Indeed, youths identified cases in which the filters did not work on dark-skinned individuals and others in which people were misgendered (see Figure 1). While some youth argued this was the filters' fault, others expressed that filters worked equally for everyone and that the problem was with the users. While evaluating the filters, youths also came up with ideas to explain how the filters worked and why sometimes they did not work on all people. To construct their explanations youths tested different aspects of the filter across faces of different people and made connections to their lived experiences. In terms of what to do about the identified biases, youths voiced different opinions with some arguing there is nothing that can or should be done, others proposing technical solutions and some considering how the filter may affect and harm real people. Some youths voiced that the filters are not problematic even after identifying and experiencing biases along race and gender. Several participants expressed that there are technical solutions (e.g., having a diverse dataset) that could be taken to incrementally improve the filters. A few youths argued that grappling with the issues they uncovered goes beyond technical solutions, for instance including regulations and addressing harm.

**Figure 1.**
Outputs generated by filter when tested by participants.

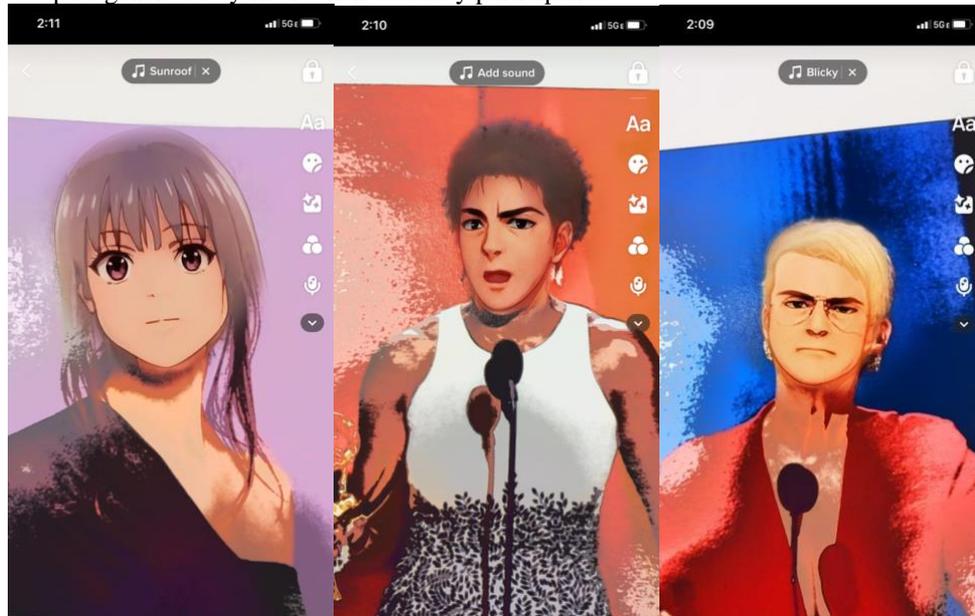

Participants tested the anime TikTok filter on Jenna Ortega, Viola Davis, and Jaime Lee Curtis identifying racial and gender biases. These included the misgendering of Curtis and Davis and the lightening of Davis' skin.

These findings expand earlier work on everyday audits conducted by adult users (DeVos et al., 2022) showing that young people also can potentially engage in productively examining how algorithmic systems work and evaluating them. This work has implications for thinking about how to promote AI/ML literacy and how to design activities that build on youth's diverse everyday understandings and that can support them in engaging with issues of algorithmic justice.

## Exploring TikTok data to understand learning practices
Ha Nguyen

TikTok supports creative expressions with multimedia affordances for users to layer text, images, videos, and multiple narratives within the same videos. Users can reuse and transform soundtracks, hashtags, and filters, potentially contributing to a community-constructed, platform-wide discourse (Zulli & Zulli, 2022). These affordances are generative to understand how users (1) adopt and augment platform trends, and (2) build on personal experiences to communicate to broad audiences and construct knowledge (Nguyen & Diederich, 2023; Nguyen & Parameswaran, 2023). In this presentation, we discuss the different layers of analyses that can shed light on users' engagement in TikTok platform practices toward meaningful learning. We focus on three layers: content analysis, hashtag analysis, and unfolding knowledge construction in the comments on TikTok videos.

## Content analysis of videos

Analyses of TikTok videos' content and video features (e.g., audio, footage, filters) reveal insights into individuals' creative expressions. We have used this approach to examine how individuals participate in critical data literacies on TikTok and meaningfully advance discourse on environmental actions (Nguyen & Parameswaran, 2023). Critical data literacies can be defined as the ability to interpret data and communicate about data from humanistic lenses (Louie, 2022; Pangrazio & Selwyn, 2019). This form of literacy examines the completeness and hierarchies behind large-scale data, while valuing small data such as personal narratives (Calabrese Barton et al., 2021; D'Ignazio & Klein, 2020). TikTok provides a platform for engaging in critical data literacies. Content creators adopt the platform's multimedia features to compile different data forms, situate large-scale climate news data in local and personal contexts, and position their experiences as data (Nguyen & Parameswaran, 2023). To illustrate, we found several instances of content creators lip-synching to the soundtrack "busy doin hot girl sh*t – Chelsea" while showing footage of themselves participating in environmental sustainability practices. The creators were making use of a viral soundtrack (and the platform algorithm) to boost the reach of their advocacy messages.

## Hashtag analysis

Hashtags (user-generated tags to enable referencing content by topics) provide another venue to examine how TikTok users show awareness of and leverage platform algorithms. Hashtags can serve as threads linking social media artifacts and communities (Yang et al., 2012). For example, in constructing a network of hashtags from public TikTok videos associated with climate understanding and action, we found several connected topics regarding climate awareness, climate change's impact, individual action, and activism (Nguyen, 2023). Findings also highlight how content creators intentionally included hashtags like #viral or #foryou to increase the content's popularity on the platform, as one way to make use of the algorithmic behaviors.

## Knowledge construction in comments

Comments on videos can also serve as productive sites to examine engagement with platform content. We have examined TikTok comments within the context of science education. We found that the site can facilitate individuals to share ideas, negotiate understanding, and explore idea dissonance (Nguyen & Diederich, 2023). Linking the types of knowledge exchange in comments and video characteristics (through content analysis) reveals the importance of original audio and relatable content for audiences to build arguments around. This content illustrates the interconnectedness of content creation and consumption, knowledge construction, and platform features.

Together, these analyses present different approaches to position TikTok as a space for individuals to incorporate multimedia features and platform algorithms into creative expressions and meaning making.

## **Teachers' reflections on TikTok for youth construction of data literacies**
Kayla DesPortes, Ralph Vacca, Camillia Matuk, Megan Silander, Anna Amato, Peter Woods, Francisco Castro, Mia Shaw

Middle school and high school youth consume and produce data across a variety of social media platforms that make heavy use of short form video (Abbas et al., 2022; Calabrese Barton, 2021). As consumers, youth use these platforms to engage with content creators who bring small data, like personal narratives, and big data, like statistics on a national sample, together into dialogue (Calabrese Barton, 2021), making it essential that they are supported in becoming critical consumers of information from these platforms. As producers, content creators already engage in critical data practices around social issues like climate action and national conflicts as they construct narratives that curate and situate data (Abbas et al., 2022; Nguyen & Parameswaran, 2023). Teachers can leverage these culturally situated platforms to provide learners with opportunities to practice critical data literacies in ways that further develop the skills they are already building outside of school.

In this work, we engaged six teachers across math, social studies, and english language arts—2 high school and 4 middle school—in individual, semi-structured interviews that prompted them to reflect on potential critical data literacy learning activities around short form video platforms like TikTok. In these interviews, we asked teachers to review and reflect upon a set of data stories from various content creators on both TikTok and Instagram, which we curated to showcase a range of ways that content creators use data, integrate personal reflections, and construct narratives for different rhetorical purposes. We then asked teachers to discuss their views around the potential of student-constructed *data stories*, specifically how they envisioned a critical data storytelling project within their existing disciplinary learning objectives.

Analysis of these interviews identified four main insights about teachers' perceptions of platforms like TikTok. First, teachers discussed methods that support learners' criticality, including having a safe space to practice, learning to investigate both the data and the content creators themselves, and developing skills to distinguish between the types and fidelity of information presented. Teachers wanted to guide their students in critiquing ideas rather than people, but some feared this would be challenging because of how the videos promoted personal and emotionally evocative narratives. Second, teachers discussed supporting argumentation and storytelling through leveraging the affordances of social media platforms, which highlighted how such platforms offer avenues for various kinds of engagement, from reaction videos in which students might identify perspectives and emotions around a data story, to tools for students to construct their own data stories. Third, teachers emphasized the potential of these platforms in building learners' abilities to humanize data. Teachers identified prior struggles to get learners to see the humans behind the numbers, and recognized the visual, narrative, and audio storytelling affordances of these platforms for communicating about data in relation to their local contexts and experiences. Lastly, teachers envisioned ways in which these methods could support learners' identity development, including their disciplinary and sociopolitical identities, by offering students opportunities for them to participate in real-world civic discourses and applying mathematical skills to these situations. However, teachers recognized the difficulty for students to articulate different perspectives than their peers, especially on social and political topics that are closely intertwined with their identities. They noted that the public nature of these platforms might intimidate students who are still figuring out how they feel about such topics. Implications of this study revealed how teachers' perspectives provide insight into how to create scaffolded opportunities for youth to engage in data practices with TikTok that build on their current activities, while also being mindful of the challenges that these platforms pose, and that merit further research.

## How to Support Critical AI literacy in K-12 Education: Raising Critical Consciousness towards Ethical AI

Selin Akgun and Christine Greenhow

Artificial Intelligence (AI) has brought drastic change to K-12 education through the integration of various AI systems, such as personalized learning, automated assessment systems, and a variety of social media platforms (Druga et al., 2018; Su & Yang, 2022). These systems keep changing the equation and modes of learning and teaching by increasingly being deployed in various educational contexts. They have their affordances in education by giving students individualized, timely, and detailed feedback; reducing teachers' workload, and supporting students to find future career paths.

Even though AI holds considerable promise in the education context, scholarly and public discussion raises critical questions about ethics and the power imbalances of AI systems (Benjamin, 2019; Holstein & Doroudi, 2021). These concerns stem from how AI learns from existing historical data without necessarily accounting for any inherent biases in the data training sets, which occasionally perpetuate inequalities and injustices, mostly towards underserved and marginalized groups (Crawford, 2021; Eubanks, 2017). The power imbalances amplify ethical and societal challenges surrounding AI use in terms of privacy, autonomy, surveillance, bias, and discrimination (Akgun & Greenhow, 2021).

**Figure 2.**
Potential Ethical and Societal Challenges Surrounding AI Use (Akgun & Greenhow, p.4).

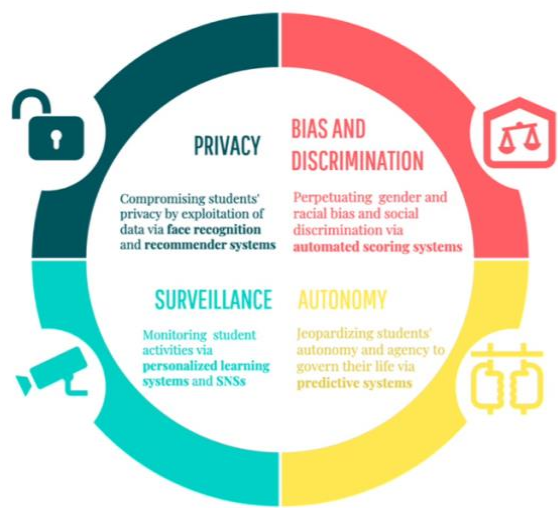

Therefore, critical discussion on how, what, and for whom these AI systems are designed is vital to consider in education (Akgun & Greenhow, 2021; Vakil et al., 2020). To build a future generation where a diverse, critical, and inclusive citizenry can participate in the use and design of future AI, we have to provide opportunities for K-12 teachers and students to develop critical AI literacy by foregrounding its ethical dimension (see Figure 2). With that goal, we aim to discuss the ways of supporting youth to consider and critique the issues of algorithmic justice and how they can become the agents of systemic change considering the ethical and societal implications of AI for their community.

In doing that, we will discuss what pedagogical approaches and social media platforms can be used to raise critical consciousness about core concepts and macro-ethical issues of AI in education. For instance, using generative AI tools and social media platforms like ChatGPT and TikTok as a part of instruction becomes critical and timely to support youth in using personally relevant and meaningful ways to make sense of data, algorithms, and issues of justice (Heaven, 2023; Zhai; 2023). As a tool, ChatGPT can be a powerful teaching and learning scaffold in the case that educators navigate it within a critical mindset and approach. In other words, teachers can guide students not only on how to just get responses and find solutions to their questions and problems, but also to support using the tool as a guide to write better (especially English language learners), research better, find relevant learning sources better, and what information to trust and criticize better (Roose, 2023). Similarly, TikTok can offer multimodal ways for students to express themselves and reflect on their creativity and thinking using various texts, images, videos, and narratives. It also can be used as an instructional tool to support youth in critically inquiring how digital tools work, how they can reflect on the issues of algorithmic justice, and use their understanding of the application to advocate for systemic change.

Through this session, we aim to contribute to the growing conversation on AI, ethics, and education and the ways of bringing and navigating these critical conversations in K-12 education. Addressing the future directions and providing recommendations for teacher educators and practitioners of AI and education is critical.

## Acknowledgments

Morales-Navarro and Kafai thank Eric Yang and Meghan Shah for support with data collection and Danaë Metaxa for their feedback. Morales-Navarro and Kafai's contribution was supported by National Science Foundation grant #2333469. Any opinions, findings, and conclusions or recommendations expressed in this paper are those of the authors and do not necessarily reflect the views of NSF, the University of Pennsylvania, Utah State University, New York University, Fordham University, Michigan State University, University of Nottingham or Stanford University.

## References

Abbas, L., Fahmy, S. S., Ayad, S., Ibrahim, M., & Ali, A. H. (2022). TikTok intifada: Analyzing social media activism among youth. *Online media and global communication*, *1*(2), 287-314.
Bhandari, A., & Bimo, S. (2022). Why's everyone on TikTok now? The algorithmized self and the future of self-making on social media. Social Media+ Society, 8(1), 20563051221086241.

Benjamin, R. (2019). Race after technology: Abolitionist tools for the new jim code. *Social forces*.
Boyd, D. (2014). It's complicated: The social lives of networked teens. Yale University Press.
Braun, V., & Clarke, V. (2012). Thematic analysis. American Psychological Association.
Burnell, K., Kurup, A. R., & Underwood, M. K. (2022). Snapchat lenses and body image concerns. New Media & Society, 24(9), 2088-2106.
Birhane, A. (2021). Algorithmic injustice: a relational ethics approach. Patterns, 2(2), 100205.
Calabrese Barton, A., Greenberg, D., Turner, C., Riter, D., Perez, M., Tasker, T., Jones, D., Herrenkohl, L.R., & Davis, E. A. (2021). Youth critical data practices in the COVID-19 multipandemic. *Aera Open*, 7, 23328584211041631.
Charisi, V., Imai, T., Rinta, T., Nakhayenze, J. M., & Gomez, R. (2021, June). Exploring the Concept of Fairness in Everyday, Imaginary and Robot Scenarios: A Cross-Cultural Study With Children in Japan and Uganda. In Interaction Design and Children (pp. 532-536).
Crawford, K. (2021). *Atlas of AI*. Yale University Press.
DeVos, A., Dhabalia, A., Shen, H., Holstein, K., & Eslami, M. (2022, April). Toward User-Driven Algorithm Auditing: Investigating users' strategies for uncovering harmful algorithmic behavior. In Proceedings of the 2022 CHI Conference on Human Factors in Computing Systems (pp. 1-19).
D'Ignazio & Klein, 2020
Druga, S., Williams, R., Park, H. W., & Breazeal, C. (2018, June). How smart are the smart toys? Children and parents' agent interaction and intelligence attribution. In *Proceedings of the 17th ACM Conference on Interaction Design and Children* (pp. 231-240).
Eubanks, V. (2017). *Automating inequality.* St Martins.
Greenhow, C., Galvin, S. M., & Staudt Willet, K. B. (2019). What should be the role of social media in education?. Policy Insights from the Behavioral and Brain Sciences, 6(2), 178-185.
Heaven, W. D. (2023, April 6). *ChatGPT is going to change education, not destroy it.* https://www.technologyreview.com/2023/04/06/1071059/chatgpt-change-not-destroy-education-openai
Holstein, K., & Doroudi, S. (2021). Equity and artificial intelligence in education: Will "AIEd" amplify or alleviate inequities in education? *arXiv preprint* . arXiv:2104.12920.
Jain, N., Olmo, A., Sengupta, S., Manikonda, L., & Kambhampati, S. (2022). Imperfect ImaGANation: Implications of GANs exacerbating biases on facial data augmentation and snapchat face lenses. Artificial Intelligence, 304, 103652.
Jenkins, H., & Ito, M. (2015). Participatory culture in a networked era: A conversation on youth, learning, commerce, and politics. John Wiley & Sons.
Kafai, Y. B., & Proctor, C. (2022). A Revaluation of Computational Thinking in K–12 Education: Moving Toward Computational Literacies. *Educational Researcher, 51*(2), 146-151.
Karizat, N., Delmonaco, D., Eslami, M., & Andalibi, N. (2021). Algorithmic folk theories and identity: How TikTok users co-produce Knowledge of identity and engage in algorithmic resistance. Proceedings of the ACM on Human-Computer Interaction, 5(CSCW2), 1-44.
Literat, I., & Kligler-Vilenchik, N. (2023). TikTok as a Key Platform for Youth Political Expression: Reflecting on the Opportunities and Stakes Involved. Social Media+ Society, 9(1), 20563051231157595.
Louie, J. (2022). Critical data literacy: Creating a more just world with data.
Metaxa, D., Park, J. S., Robertson, R. E., Karahalios, K., Wilson, C., Hancock, J., & Sandvig, C. (2021). Auditing algorithms: Understanding algorithmic systems from the outside in. Foundations and Trends® in Human–Computer Interaction, 14(4), 272-344.
Morales-Navarro, L., & Kafai, Y. B. (2023). Conceptualizing Approaches to Critical Computing Education: Inquiry, Design, and Reimagination. In Past, Present and Future of Computing Education Research: A Global Perspective (pp. 521-538). Cham: Springer International Publishing.
Nguyen, H. (2023, March). TikTok as Learning Analytics Data: Framing Climate Change and Data Practices. In *LAK23: 13th International Learning Analytics and Knowledge Conference* (pp. 33-43).
Nguyen, H., & Diederich, M. (2023). Facilitating knowledge construction in informal learning: A study of TikTok scientific, educational videos. *Computers & Education*, *205*, 104896.

Nguyen, H., & Parameswaran, P. (2023). Meaning making and relatedness: exploring critical data literacies on social media. *Information and Learning Sciences*.

Pangrazio, L., & Selwyn, N. (2019). 'Personal data literacies': A critical literacies approach to enhancing understandings of personal digital data. *New media & society*, *21*(2), 419-437.

Pew Research Center, (2022). *Teens, Social Media and Technology 2022*. Pew Research.

Roose, K. (2023, January 12). *Don't ban ChatGPT in schools. Teach with it.* https://www.nytimes.com/2023/01/12/technology/chatgpt-schools-teachers.html

Shen, H., DeVos, A., Eslami, M., & Holstein, K. (2021). Everyday algorithm auditing: Understanding the power of everyday users in surfacing harmful algorithmic behaviors. Proceedings of the ACM on Human-Computer Interaction, 5(CSCW2), 1-29.

Su, J., & Yang, W. (2022). Artificial intelligence in early childhood education: A scoping review. *Computers and Education: Artificial Intelligence*, *3*, 100049.

Taylor, S. H., & Choi, M. (2022). An Initial Conceptualization of Algorithm Responsiveness: Comparing Perceptions of Algorithms Across Social Media Platforms. Social Media+ Society, 8(4), 20563051221144322.

Vakil, S., Marshall, J., & Ibrahimovic, S. (2020). "That's Bogus as Hell!": Getting Under the Hood of Surveillance Technologies in an Out of School STEM Learning Environment.

Vakil, S., & McKinney de Royston, M. (2022). Youth as philosophers of technology. Mind, Culture, and Activity, 1-20.

Yang, L., Sun, T., Zhang, M., & Mei, Q. (2012, April). We know what@ you# tag: does the dual role affect hashtag adoption?. In *Proceedings of the 21st international conference on World Wide Web* (pp. 261-270).

Walker, R., Sherif, E., Breazeal, C.: Liberatory computing education for African American students. In: 2022 IEEE Conference on Research in Equitable and Sustained Participation in Engineering, Computing, and Technology (RESPECT), pp. 85–89. IEEE (2022)

Zhai, X. (2023). ChatGPT for next generation science learning. XRDS: Crossroads. The ACM Magazine for Students, 29, 42–46. https://doi.org/10.1145/3589649

Zulli, D., & Zulli, D. J. (2022). Extending the Internet meme: Conceptualizing technological mimesis and imitation publics on the TikTok platform. *New media & society*, *24*(8), 1872-1890.